# Summary of the ACAT Round Table Discussion: Open-source, knowledge sharing and scientific collaboration

Federico Carminati

CERN-PH, 1211 Geneva 23, Switzerland

E-mail: Federico.Carminati@cern.ch

Denis Perret-Gallix

LAPP/IN2P3, CNRS, 74941 Annecy-le-Vieux, France

E-mail: Denis.Perret-Gallix@in2p3.fr

Tord Riemann

DESY, 15738 Zeuthen, Germany

E-mail: Tord.Riemann@desy.de

**Abstract.** Round table discussions are in the tradition of ACAT. This year's plenary round table discussion was devoted to questions related to the use of scientific software in High Energy Physics and beyond. The 90 minutes of discussion were lively, and quite a lot of diverse opinions were spelled out. Although the discussion was, in part, controversial, the participants agreed unanimously on several basic issues in software sharing:

- The importance of having various licensing models in academic research;
- The basic value of proper recognition and attribution of intellectual property, including scientific software;
- The user respect for the conditions of use, including licence statements, as formulated by the author.

The need of a similar discussion on the issues of data sharing was emphasized and it was recommended to cover this subject at the conference round table discussion of next ACAT. In this contribution, we summarise selected topics that were covered in the introductory talks and in the following discussion.

## 1. Introduction

ACAT, the "Workshop on Advanced Computing and Analysis Techniques in Physics Research", is with us since more than 20 years now. It began with a software workshop in Lyon in 1990, went on under the name AIHENP since 1992 in La Londe Les Maures, and the name ACAT was first coined for the meeting at Fermilab in 2000. Following this nomenclature, we have now the 15th ACAT, see [1]. High Energy Physics is at the heart of ACAT, which traditionally includes three tracks: "Computing technology for physics research", "Data analysis - algorithms and tools", and "Techniques and methods in theoretical physics".

At ACAT 2008, there was a plenary talk on "Aspects of intellectual property law for HEP software developers" [2]; see also [3]. At ACAT 2009, we had a round table discussion devoted to "Event generation – are we ready for LHC?" [4]. In one section, "The quest for publicly available code" was discussed. In fact, software intellectual property related questions are getting more and more important in our research activity

This time, at ACAT 2013, we discussed "High Energy Physics software, knowledge sharing and scientific collaboration" not from the point of view of law, strongly affected by national regulations but, our field being vastly international, focusing on ethics and fairness of use There is a lot of literature available on the subject, but the complexity of legal regulations makes it of little practical help; see several Wikipedia articles on the topics, e.g. [5].

During the round table we specifically addressed the problems related to software sharing when "Good Scientific Practice" refers, in general, to "data, text or theories". For a summary on good practice see e.g. "Springer Policy on Publishing Integrity - Guidelines for Journal Editors" [6].

In recent decades, developing scientific software has become a full-fledged scientific activity of its own, see e.g. the discussions in [7, 8], or the famous CERNLIB history [9] with several licence regulations [10] for its different parts, see also the historical overview on the ZFITTER package [11] with a CPC licence, and many other long-term software projects. The very existence of conference series like ACAT together with dedicated scientific journals like Elseviers "Computer Physics Communications" and its companion "Computer Physics Communications Program Library - Programs in Physics & Physical Chemistry" at Belfast University, issuing the well- known CPC software licence statements [12], are clear signs of this evolution.

From our scientific experience, we have learned that there are specific rules to follow when dealing with software in knowledge sharing, even if, before the conference, controversial positions were expressed on the need or usefulness of discussing these issues.

Actually, High Energy Physics and, to some extent, even theoretical particle physics has become more and more a kind of industry. Due to this trend, the contacts between cooperating scientists tend to become more anonymous and virtually driven by Internet search. At the same time, the competition has become harsher and less open.

Knowledge sharing in early times was simple:

- Typically an author wrote software for his/her project to get numerical results. Sometimes he/she could hand over the Fortran code to a more or less close colleague. Sometimes the name of the author and a date was simply written in the code, sometimes not even that. Often people just copied what they needed into their Fortran codes without specific mention of the origin.[1]
- The other extreme was the way CERNLIB [9] was managed; its software was nearly free but better protected; see the 2004 licence statement for a typical function in CERNLIB [15].

Even if no rules or "conditions of use" were formulated, there was a common consensus on how to cite the software used in the scientific publication. Today we call this "proper attribution".

Actually, knowledge sharing today is much more complex than in the old days:

- There are many means of distribution, most of them are more or less anonymous.
- The software is now, primarily, written with the purpose of being used by other people: experimentalists or theorists and, not only, by the original authors.

---

[1] We again like to use as an example the ZFITTER project [13]. The package exists since 1985, and is distributed among users since 1992. For the first time, in the 1999 version of its description [14], the software pieces written by others were listed completely.

- The complexity of the software becomes higher by orders of magnitude, and it often results in many man-years involvement. This has consequences. The main one, for our discussion here, is the fact that software itself is the final product of sometimes highly complex scientific studies, and not only a by-product of a research activity.

Facing to these difficulties, people react differently, and during the round-table, we had the following reactions:

- When I create software, I want to get cited for its use in a way which I define. Sometimes this is achieved by applying a GPL-type licence (see http://en.wikipedia.org/wiki/GNU_General-_Public_License), sometimes by references to published articles.
- Our software has not to be touched by the user, because we guarantee for its high standards or because it is a standard candle [an etalon] for others, and it was created as such. So, please link my software to yours, or refrain from using it. Or, please write your interface to my package as a whole. Or, please contact me directly.

The following rather unethical statements, some of them often heard, express contradicting opinions of the use of software:

- Any software on Internet with anonymous download is, for me, open-source software.
- And I define by myself what the term "open-source" means: just free for any use.
- I need open-source software of this type because I have to adapt and develop your software for my purposes, in the interest of scientific progress.
- If you do not like misuse, you should not publish your software on Internet. This is the only safe way. I myself go this way.
- As there are no commonly accepted rules for the use and citation of open-source software, of course, I need not cite the software when using it. Why should I? And I can do what I like with the software I got, even if the authors claim to have a "licence".

We do not have the space to go into details of the discussion on definitions of "open-source software" or "licence". Two positions were expressed, though, in the discussion:

- It is the contradiction of the opinions, all with some justifications, which made the round table discussion lively. We discussed mainly what one might call source-open software, to mark the fact that the source code is open for read-only. While, the word open-source refers to some specific conditions of use. To get a flavour of the current confusion, one may consult the various articles in various national editions of Wikipedia.
- We agreed that the software licences mentioned here are not licences in the usual, legal meaning. They are, for us, standardized formulations of "conditions of use", enforced by authors or by their organizations. This approach reflects scientific practices and makes us independent of national law - as long as the rules are followed.

We discussed exclusively about academic, basic research, and in particular only about academic software, to be specific. This excluded e.g. any reference to commercial software.

- We live in an international community. As a consequence, national law, national licences, institutional regulations are not automatically valid. See the Wikipedia article on the Bern convention for the protection of literary and scientific works [16].
- We researchers at ACAT work on long-term projects, often in teams, sometimes really huge teams of several thousand people (e.g. ATLAS and CMS collaborations), and, in addition, with a variable membership. As a consequence, the definition of an "author" as well as his practical rights in a project are difficult to define with precision.

- Academic researchers career and funding depend, in many respects, on the recognition of the contributions to scientific progress. And this recognition has influence on:
  - the project budget
  - the resources that can be used, e.g. large disc spaces and clusters of computers for large-scale calculations
  - the possibilities to hire PhD and postdoc.
  - the permanent position that can be obtained
  - the recognition in the science "hall of fame" [non-monetary recognition]

For all these reasons, since the "Renaissance" the tradition of citation - or attribution - of the work of others [works = creations] became more and more an essential part of scientific ethics in European and world-wide basic research.[2]

Attribution applies either because the work relies on others' work, or because the work of others is directly integrated. This need for "attribution" in basic research is beyond commercial arguments, but not beyond material interests.

The usual mean of scientific attribution in basic academic research is citation as it is usually done in well-written articles.[3] But the balance between Competition and Cooperation gets destroyed when researchers use others' achievements without attribution.

In practice, there are additional expectations of the "creators of scientific work" toward their users that are well accepted by the society (in Germany, they are called "personal copyrights"). See for example, the regulations in the Internet for photographs, videos, music, texts, etc. It is well known by everyone that one has to care about the "conditions of use" when downloading anything.

For software, there may be certain very specific regulations. E.g. in German law this applies.[4]

Because national law often cannot be applied in practice, it is of high importance that researchers feel an ethical need to respect the conditions of use of any work, formulated by the authors of software. Licences are integral part of these "conditions of work usage".

Examples of popular licence statements:

- Gnu General Public licence = GPL and the derivatives, like e.g. the lesserGPL [17].
  It is often used and/or recommended. But it is not quite appropriate for academic software because it does not enforce proper citation.
- Creative Commons Licences = CC licences, with a complete tree of derivatives [19].
  This type of licences seems to be appropriate for scientific work, because it defines how to deal with attributions. Nevertheless, it is not often recommended for software licencing.
- Computer Physics Communications software deposit licence [12].
  This licence is in use since decades by many high energy physics software packages, e.g. GEANT, MINUIT, ZFITTER, FF, etc. Since recently, the users of the CPC software deposit have a choice of licence.
- FLUKA licence [20].
  The licence is rather strict.
- POWHEG BOX licence [21, 22].
  Example of guidelines that you may find short, simple and useful.

---

[2] One of the first holders of a kind of licence police, released by the emperor Maximilian, was the German painter Albrecht Dürer (1471-1528). The perhaps most famous scientific struggle about copyright was put forward by Issac Newton (1642-1727) and Gottfried Wilhelm Leibniz (1646-1716).

[3] The licence GPL [17] does not demand attribution in an article based on the use of GPL-licenced software. For this reason, the GPL is not truly appropriate for scientific purposes.

[4] For German speaking colleagues: German copyright law, see [18]; it is easy to read and understand. Although, some practiced regulations of that law deviate heavily from a naive understanding of the text.

- MCnet guidelines [23].
  They recommend GPLv2.
- HepForge, a development environment for High Energy Physics software projects [22, 24].
  No licence statement has been released.

These examples are shown in order to demonstrate one conclusion from the discussion: There is not a single widely appropriate licence model for all academic software projects and their purposes and we must keep at hand a variety of licence agreements fitting each one particular needs.

Finally, the decision to select a licence is taken jointly by the authors of the software, the employer and the funding agencies. The last two, the "third parties" often will not get involved explicitly. But, in other cases, like for the huge competing collaborations (ATLAS, ALICE, or CMS), the "licencing rules" can get much stronger and it might happen that the usage of software be restricted, exclusively, to the collaboration members.

In the view of some experts of law, the software licences of High Energy Physics are "so-called licences", i.e. no true licences, because domestic law might have different definitions on what is a licence. But, rejecting these licences is not a good approach as they are, at least for the best of them, a commonly accepted frame for the use of others' work. They are, in practice, the rules of the game.

Finally, the world of research is largely embedded in the society. Researchers in academic basic research are supported by the society and, as such, they do their best to:

- Show honesty in their research;
- Make the scientific achievements publicly available.

To summarize:

- We, software developers and authors are taking an active part in the design of the ethical regulations on the creation of knowledge.
- Conditions of use, including licences, are formulated by the authors, their employers and the funding agencies. Every efforts should be made to have them as convenient as possible to the
  users.
  Publish whatever work you have been producing: ASAP = As Source-open As Possible.
- Institutions should have compliance statements. This helps to regulate the relations with the researchers, in case of conflicting situations.
  Definitions of plagiarism, ombudsperson, and lists of sanctions are perhaps also needed, but the best approach is to prevent conflicts.

The summary of the Round Table Session on the last day's final session made evident that there is a broad understanding on the problems related to the collaborative use of software in the community. From time to time, problematic situations may arise, but it is usually clear how to handle them.

Concerning the collaborative handling of data, the situation is less clear. This is mainly due to the fact that such a large amount of data as produced by the LHC collaborations has never been collected in the past, so that there is less experience.

It might be a good idea to devote a round table discussion at the next ACAT conference to the problems of data sharing, data security, data property and data long term integrity. It would also fit well into this year's round table title:

Open-source, knowledge sharing and scientific collaboration